\documentclass[reqno]{amsart}       				
\usepackage{etex}

\usepackage{amsmath,amsthm,amssymb,stmaryrd,latexsym,mathrsfs}
\usepackage[usenames,dvipsnames]{color}			
\usepackage{graphicx,epic}            			
\usepackage{pstricks,pst-tree,pstricks-add}		
\usepackage{hyperref}                   		
\hypersetup{									
	colorlinks=true,      		             	
	linkcolor=Blue,                        		
	filecolor=Blue,                     		
	citecolor=Blue,    	        				
	urlcolor=Blue,                 	   		 	
	bookmarks=true,                    			
	bookmarksopen=true,           				
	pdffitwindow=false,               	 		
	pdfstartview={FitH},               			
	pdfauthor={Yang Liu},           			
}
\usepackage{setspace}                   		
\usepackage{framed}                             
\usepackage[arrow,arc,all]{xy}              	
\usepackage{multicol,multirow,array,ctable}    	
\usepackage{threeparttable}						
\usepackage{enumerate}             				
\usepackage{marginnote}               			
\usepackage{natbib,hypernat}                    
\usepackage{comment}							
\usepackage{pdflscape}							

\theoremstyle{definition}
\newtheorem{thm}{Theorem}

\newtheorem{cor}[thm]{Corollary}

\newtheorem{defn}[thm]{Definition}

\newtheorem{prop}[thm]{Proposition}

\theoremstyle{definition}

\theoremstyle{remark}

\newcommand{\corref}[1]{Corollary~\ref{#1}}
\newcommand{\defnref}[1]{Definition~\ref{#1}}

\newcommand{\figref}[1]{Figure~\ref{#1}}

\newcommand{\lemref}[1]{Lemma~\ref{#1}}
\newcommand{\propref}[1]{Proposition~\ref{#1}}

\newcommand{\tableref}[1]{Table~\ref{#1}}

\newcommand{\thmref}[1]{Theorem~\ref{#1}}

\def\o{{\omega}}

\def\x{{\xi}}

\def\A{{\mathcal A}}    	    			
\def\B{{\mathcal B}}

    	   		\def\FF{{\mathscr F}}

\def\I{{\mathcal I}}    			
    			
\def\K{{\mathcal K}}

\def\P{{\mathcal P}}

\def\T{{\mathcal T}}

\def\Real{\mathbb{R}}

\def\xn #1{{#1}_1,\ldots,{#1}_n}

\def\x0{x_0,\ldots,x_{n-1}}

\def\op #1{\ensuremath{\textrm{\normalfont #1}}}

\def\TC{\op{TC}}

\def\brkt #1{\langle #1 \rangle}

\long\def\sfootnote[#1]#2
{\begingroup\def\thefootnote{\fnsymbol{footnote}}
\footnote[#1]{#2}\endgroup}                                         

\usepackage{mathpazo, helvet, courier}				
 
\def\bbrkt#1{\llbracket #1\rrbracket}
\def\Prio{{{\Pr}^{i}}_\o}
\def\blind{\B^i(\Omega)}
\begin{document}
\title{Towards a Unified Belief Structure in Games with indeterminate probabilities} 
\author{Yang Liu}
\address{Department of Philosophy, Columbia University, New York, NY 10027}
\email{yl2435@columbia.edu}
\keywords{representing  beliefs, indeterminate probabilities, Bayesianism, blindspot, sure-thing principle, ambiguity.}
\thanks{ 
Thanks are due to Haim Gaifman, Joe Halpern, Issac Levi,  Rohit Parikh, and Andr\'{e}s Perea for inspiring exchanges, and to the audience at the 4th Formal Epistemology Festival (Konstanz 2012) and the 67th European meeting of the Econometric Society (EEA$|$ESEM 2013) for helpful comments where preliminary versions of different parts of this paper were presented.}

\maketitle
\begin{abstract} 
This paper provides an  analysis of different formal representations of beliefs in epistemic game theory.  The aim is to attempt a synthesis of different structures of beliefs  in the presence of {\it indeterminate probabilities}. Special attention is also paid to the decision-theoretic principle known as the  {\it thesis of no subjective probability for self-action}. Conditions in cope with this principle are given which underlie the interrelationships between different models of beliefs, and it is shown that under these conditions different doxastic structures can be coherently unified.
\end{abstract}


\section{Introduction} 
 
The epistemic approach to non-cooperative game theory places in the center the analyses of {\it epistemic  mutual expectations} of different parties of a game, where the players' choices of the best courses of action are determined not only by their beliefs about the structure of the game (including players, actions, payoffs, and preferences) but also by their beliefs of the beliefs of the other players, and so on. A mathematical model that emphasizes the epistemic dimension of a game hence needs to incorporate, as a constituent component of the model, a systematic representation of players' beliefs. Two main types of formal representation are widely adopted in the philosophical literature, namely, the {\bf Kripke doxastic structure $\K$}  and the {\bf Bayesian probabilistic structure $\P$}. Both structures are constructed on a {possible-world} framework where a player's belief expresses certain attitudes towards a given proposition (or an event, in the probabilistic case) which is further characterized as a set of doxastic {possibilities}. 
In a game-theoretic setting, each possibility corresponds to a complete description of an alternative way the game may evolve, and it is customary in game theory that players' propositional attitudes be subject to informational analysis using {\bf information structures $\I$}, where different representations of beliefs are seen as different measures of information distributions among the players. Another type of doxastic representation that has a distinctive place in game-theoretic analyses is the use of various {\bf Harsanyi type structures $\T$}. Depending on the particular modeling, a type structure is a probability function that measures a player's uncertainties about other players' actions (or/and payoffs, beliefs, etc.), which is considered as an integral part of  overall doxastic assessments of the player.   

One of the main aims of the present paper is to provide a synthesis of the aforementioned belief structures ($\K,\I,\P,\T$). Numerous discussions on different representations of beliefs in games and their interrelationships are readily available in, for instance, \cite{Battigalli99}, \cite{Halpern03c},  \cite{Brandenburger07}. The current presentation is distinctive in that, in all Bayesian frameworks discussed below, it is taken as a fundamental assumption that the players' beliefs be measured by {\bf indeterminate  probabilities} (sets of subjective probabilities) instead of sharp/determinate ones, an approach motivated by the philosophical thesis that the agent's credal state shall be represented by indeterminate credences. The literature on indeterminate probabilities is vast, see, for instance, \cite{Levi74}, \citet{Walley91}, \cite{Gilboa89} and \cite{Seidenfeld95}. In this paper, we do not engage in the philosophical debate on whether or not rational agents always have sharp (subjective) probabilities,  rather, it is taken for granted that players' credal states be represented by sets of probabilities. Focus is however placed on investigating as to how different belief structures fit in to one another with the idea of indeterminate probabilities in sight.

One concept that plays an important role in the discussions of belief structures below  is that of {\it doxastic blindspots}.  Roughly speaking, in a  possible-world framework equipped with accessibility relations, a doxastic blindspot is taken to be a state/world that is not considered by the decision maker as doxastically possible. We show, through the analysis of the impossibility theorem of \cite{Aumann76}, that the uniformity of doxastic blindspots among the players in a group is the key to the convergence of group opinion, which, however, is presumably a very strong assumption on the part of players. In discussing Bayesian probabilistic models for games, special attention are also given to the decision-theoretic principle known as the {\it thesis of no probability for self-action} defended by \cite{Savage72}, \cite{Spohn77},   \cite{Levi89, Levi96}, \cite{Gaifman99},  \cite{Liu13} among others. That is, 
\begin{quotation}
	no subjective/personal probabilities should be assigned by the players to {\it their own future actions}.
\end{quotation}
This principle, as we shall discuss in greater detail in Section \ref{sec:remarks:09-03-13},  posits a challenge to the epistemic configurations of many of the existing game-theoretic models. In coping with this principle we make clear distinction, in the interpretation of the formalism developed below, between probability assignments assigned globally to state-descriptions of a given game situation (a third-person perspective) and those assignments that characterize players' first-person subjective assessments of the unfolding of the underlying game.%

The plan of the paper is as follows: since information takes the center stage in game-theoretic analyses, we first present an extended discussion on information structures and its relation to Kripke structures, where a purely information-theoretic version of \citeauthor{Aumann76}'s (\citeyear{Aumann76}) impossibility theorem is proved. The proof relies on the notion of doxastic blindspots and a generalized \citeauthor{Savage72}'s sure-thing principle ({\bf GSTP}) which will be made clear shortly. This will be followed by the introduction of different  representations of beliefs widely adopted in the game-theoretic literature. Conditions (\ref{eq:bs:info function}, \ref{eq:bs:info to leadsto}, \ref{eq:bs:blindspot}, \ref{eq:bs:type and prior}, and \ref{eq:bs:type and accessibility}) are given in the attempt to correlate different belief structures with one another, and it is shown that under these conditions there is a sense in which the belief structures can be coherently unified in epistemic games. Some conceptual issues concerning the asymmetry of epistemic viewpoints in existing game-theoretic models are discussed in Section \ref{sec:remarks:09-03-13}.
  
\section{Information Structure}\label{sec:bs:kripke}

\subsection{Kripke Structure}
Let $\Omega$ be a (finite) set which is referred to as the state space. A Kripke model over   $\Omega$ is a relational structure which is distinguished with a binary relation $\leadsto \;\subseteq \;\Omega\times \Omega$. In a  Kripke model of beliefs, $\leadsto$ is referred to as a {\it doxastic  accessibility relation} among possible states. 
Intuitively, $\o\leadsto^i\o'$ says that, from the perspective of player $i$, $\o'$ is considered doxastically possible in state $\o$. We also say that $\o'$ is \emph{$\leadsto^i$-accessible} from $\o$. The following is a list of properties of $\leadsto$ that are commonly adopted in a doxastic model: for any $\o,\o',\o''\in \Omega$, 	
\begin{description}\label{pl:sf:DB}
		\item[Seriality] for each $\o$ there exists an $\o'$ such that $\o\leadsto\o'$.
		\item[Transitivity]    if $\o\leadsto\o''$and $\o''\leadsto \o'$  then $\o\leadsto \o'$.
		\item[￼Euclid] if $\o\leadsto \o'$ and $\o\leadsto \o''$ then $\o'\leadsto \o''$.
\end{description}
\subsection{Information structures} Define function $\I^i:\Omega\to 2^{\Omega}$  by
\begin{equation}\label{eq:bs:info function}\tag{A1}
	\I^i(\o)=\{\o'\in \Omega\mid \o\leadsto^i\o'\}.
\end{equation}
Then $\I^i(\o)$ is the set of states that are $\leadsto^i$-accessible from $\o$,  call $\I^i(\o)$ the \emph{information set} of  $i$ in state $\o$, the intended interpretation is that $\I^i(\o)$ contains all the relevant information that can be accessed by $i$ at $\o$. Player $i$ is said to be \emph{informed of} certain {\it event} $E$ (a set of states) in state $\o$ if $\I^i(\o)\subseteq E$ (i.e., if the information  $i$ possesses at $\o$ is contained in $E$). 
For any $E\subseteq \Omega$, denote by $\I^i(E)$ the set of all states that are $\leadsto^i$-accessible from the states in $E$, i.e., $\I^i(E)=\bigcup_{\o\in E}\I^i(\o)$.
Further, let
	\begin{equation}\label{eq:consensu:info structure}
	\I^i=\{\I^i(\o)\mid \o\in \Omega \}
	\end{equation}	
be the \emph{information structure} of player $i$. Hence $\I^i$  provides a full description of what information player $i$ has in each state.\footnote{
		Here we introduce a systematic ambiguity using $\I^i$ to denote the information structure of player $i$ and using $\I^i(\cdot)$ to denote the information function of $i$. There shall be no danger of confusion: $\I^i(\o)$ is the information set of player $i$ at $\o$, which is also an element of $\I^i$. 
} 

Alternatively, one can take the information function $\I^i:\Omega\to 2^\Omega$ of player $i$ as primitive and define $i$'s accessibility relation $\leadsto^i$ over $\Omega$ by 
	\begin{equation}\label{eq:bs:info to leadsto}\tag{A2}
		\leadsto^i:=\Big\{(\o,\o')\in \Omega\times \Omega \mid  \o'\in \I^i(\o)\Big\}.
	\end{equation} 
Consider the following properties of an information structure $\I$: for any $\o\in \Omega$,
\begin{description}\label{pl:sf:DB}
		\item[Viability] $\I(\o)\neq\emptyset$.
		\item[Inclusion]  if $\o'\in \I(\o)$ then $\I(\o')\subseteq\I(\o)$.
		\item[Mutuality] if $\o',\o''\in \I(\o)$ then $\o''\in\I(\o')$ and $\o'\in \I(\o'')$. 
\end{description} 
The following theorem establishes a basic relation between  properties of doxastic Kripke models and that of information structures, the proof is immediate from \eqref{eq:bs:info function} and \eqref{eq:bs:info to leadsto},  and hence omitted. 
 
\thm\label{thm:ck:property structures} 
	Let $\I$ be an information structure and $\leadsto$ be the corresponding accessibility relation  then,
	\begin{enumerate}
	\item\label{thm:ck:property structures:1}  
		$\leadsto$ is serial if and only if $\I$  is  {viable};
	\item\label{thm:ck:property structures:4} 
		$\leadsto$ is transitive if and only if  $\I$  is  {inclusive};
	\item\label{thm:ck:property structures:5} 
		$\leadsto$ is Euclidean if and only if  $\I$  is  {mutual}.
	\end{enumerate}
\endprop

\defn\label{defn:bs:divisible}
	 An information structure $\I$ is said to be {\it divisible} if it is (a) viable, (b) inclusive, and (c) mutual; $\I$ is {\it partitional} if it is divisible and $ \I(\Omega)=\Omega$.  
\enddefn

\prop\label{lem:bf:divisible property} 
	 If $\I$ is divisible then, for any $\o,\o'\in \Omega$, either $\I(\o)\cap\I(\o')=\emptyset$ or $\I(\o)=\I(\o')$.
\endprop

It is easy to see that if $\I$ is divisible then, by \thmref{thm:ck:property structures}, the corresponding accessibility relation $\leadsto$ forms an equivalence relation over $\Omega$, in this case we have that $\I(\o)=[\o]_\leadsto$ for all $\o\in \Omega$.

\subsection{Doxastic blindspot} 
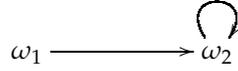
\begin{figure}[h] 
	\begin{displaymath} 
	\xymatrix{
	    {\o_1} \ar[rr] & &   {\o_2} \ar@(lu,ru)[]   
		}
	\end{displaymath} 
	\caption{State $\o_1$ is a {\it blindspot} for player $i$.}\label{fg:paradox:S4}  
\end{figure}

The discussion above enables a classification of states in $\Omega$. Note that  if $\I(\Omega) $ is a proper subset of $\Omega$ then it follows that there are some states that do not fall in any information set, or equivalently, they are not   $\leadsto^i$-accessible for $i$ from any state in $\Omega$. We refer to such states as player $i$'s {\it doxastic blindspots}. For instance, in \figref{fg:paradox:S4}, let $\Omega=\{\o_1,\o_2\}$ and $\I^i(\o_1)=\I^i(\o_2)=\{\o_2\}$, then $\o_1$ is a doxastic blindspot for $i$.\footnote{Our concept of doxastic blindspot is altered from \cite{Collins96},  who in turn introduced the term from \cite{Sorensen88}. The analysis here is a continuation of \cite{Collins96}.
}
 
\defn\label{defn:bs:blindspots}  
	For any $\o\in \Omega$, $\o$ is said to be a {\it doxastic blindspot} (or {\it blindspot} for short) of player $i$ if there is no $\nu\in \Omega$ such that $\o\in\I^i(\nu)$. The set of  all blindspots of $i$ is  denoted by $\B^i(\Omega)$.
\enddefn

\prop\label{lem:bs:Info and blind}
	$\I^i(\Omega)\cap\B^i(\Omega)=\emptyset$ and $\I^i(\Omega)\cup\B^i(\Omega)=\Omega$.
\endprop


\prop \label{lem:bs:Info and i j}
	Suppose that, for any $i,j\in N$, the information structures $\I^i$ and $\I^j$ are divisible, then 
	\begin{enumerate}
		\item\label{lem:bs:Info and i j:1}
			   $\o\notin\I^i(\o)$ if and only if $\o\in \blind$;
		\item\label{lem:bs:Info and i j:2}
			  if $\nu\in \I^i(\o)$ then $\nu\in \I^i(\nu)=\I^i(\o)$;
		\item\label{lem:bs:Info and i j:3}
			  if $\I^i(\Omega)=\I^j(\Omega)$ then, for any $\nu\in \I^i(\Omega)$,  $\nu\in \I^j(\nu)$.
	\end{enumerate}
\endprop

It shall be pointed out that the existence of blindspots  differentiates a doxastic information model from an epistemic one, the latter is widely used in representing knowledge where it is assumed that $\o\in \I^i(\omega)$ for all $\o\in \Omega$, that is, it is assumed that it is impossible that the players' information sets exclude ``the true state of the world'' (and hence there is no blindspot). The assumption is often referred to as the ``truth condition'' of information sets which echoes the substantive assumption that knowledge is infallible. Many interesting results in the game-theoretic literature hinge on this conditions (for instance, \cite{Aumann76}, \citet{Geanakoplos82}, \citet{Parikh90}, \cite{Aumann95}, \cite{Aumann98a}). 
The assumption however is not uncontroversial, especially when observing from the agent's point of view, it is unclear as to  how the existence of blindspots can be eliminated by stipulating that player's information be always truthful.   
In the next two sections, we provide a purely information-theoretic analysis of \cite{Aumann76}, we discuss a scenario under which a variant of Aumann's impossibility theorem still holds under weakened conditions. The analysis relies on (1) a notion of {\it common information}, which is analogous to the concept of common knowledge (or common beliefs), and (2) a generalized Savage's sure-thing principle.
  
\subsection{Common information}\label{sec:BS:common info}
In the interactive situation,  let $\leadsto^N$ be the smallest transitive set that contains all the $\leadsto^i$'s, that is,  
	\begin{equation}\label{eq:consensu:info function222}
	\leadsto^N\;:= \,\TC \Big(\bigcup_{i\in N} \leadsto^i \Big),  
	\end{equation}
where `TC' stands for the transitive closure operator. Then, relation $\leadsto^N$ represents the maximum reachability of all $\leadsto^i$'s (cf. \lemref{lem:info:basics PN pr}\eqref{lem:info:basics PN pr:1} below). Call $\leadsto^N$ the \emph{group accessibility relation} of $N$. For any $(\o,\o')\in\,\leadsto^N$, we also say  that $\o'$ is $\leadsto^N$\emph{-accessible from} $\o$.  From the group accessibility relation $\leadsto^N$  a corresponding notion of \emph{group information function} $\I^N:\Omega\to 2^{\Omega}$  can be defined by
	\begin{equation}\label{eq:consensu:info function2}
	\I^N(\o)=\{\o'\in \Omega\mid \o\leadsto^N\o'\}.
	\end{equation}
And let $\I^N$ be the \emph{group information structure} such that\footnote{We can also take players' information structures $\I^i,\ldots,\I^n$ as primitive and define  group information structure $I^N$ as the {\it meet} of the $\I^i$s, i.e., $\I^N=\bigwedge_{i\in N}\I^i$, and hence the group accessibility relation $\leadsto^N$ can be defined by 
\begin{equation}\label{eq:consensu:info function222a}\tag{\ref{eq:consensu:info function222}'}
	\leadsto^N\;:=\{(\o,\o')\in \Omega\times \Omega \mid  \o'\in \I^N(\o)\}.  
	\end{equation}
} 
	\begin{equation}\label{eq:infomodel:comon structure}
	\I^N=\{\I^N(\o)\mid \o\in \Omega\}. 
	\end{equation}
\defn\label{defn:infomodel:cevent}
	Let $E$ be any event,  say that $E$ is \emph{common information} among members of group $N$ at $\o$, if $\I^N(\o)\subseteq E$.	
\enddefn
The following lemma exemplifies some basic properties of the group accessibility relation $\leadsto^N$ and   the group information structure $\I^N$.

\prop\label{lem:info:basics PN pr}
	Let $\I^i,\leadsto^N$, and $\I^N$ be defined as above, then the following hold
	\begin{enumerate}
		\item\label{lem:info:basics PN pr:1}
			For any $(\o,\o')\in\, \leadsto^N$, there corresponds a sequence $i_1,i_2,\ldots,i_k\in N$ and a sequence of states $\o_0,\o_1,\ldots,\o_k\in \{\nu\mid \o\leadsto^N \nu\}$ with $\o_0=\o$ and $\o_k=\o'$ such that $\o_0\leadsto^{i_1}\o_1\leadsto^{i_2}\cdots\leadsto^{i_k}\o_k$, where $0\le k<\infty$.
			
		\item\label{lem:info:basics PN pr:2}
			For any $\o\in \Omega$ we have
			\begin{equation*} 
			\I^i(\o)\subseteq \I^N(\o) .
			\end{equation*}
			
		\item\label{lem:info:basics PN pr:3}
			For any $\o\in \Omega$ and for any $i\in N$, we have
			\begin{equation*} 
			 \I^i\big(\I^N(\o)\big)\subseteq \I^N(\o) .
			\end{equation*}
			
		\item\label{lem:info:basics PN pr:4}
			Given any $\o\in \Omega$, if, for any $i,j\in N$, $\I^i$ and $\I^j$ are divisible and  $\I^i(\Omega)= \I^j(\Omega),$ then
			\begin{equation*} 
			 \I^N(\o)=\I^i\big(\I^N(\o)\big).
			\end{equation*}
			
	\end{enumerate}
\endprop

\subsection{Agreeing to disagree}\label{sec:BS:common Agreeing to disagree}  
Savage proposed his ``sure-thing principle'' (STP) as one of the basic postulates in his theory of decision making under uncertainty \cite[{\bf P2} in][p.23]{Savage72}. The principle is derived from, what he calls, a ``loose'' version of STP which says that {if} a decision maker will take certain action conditional on the occurrence of some event and she will take the same action if the event does not occur, {then} she shall take the action without taking into account the occurrence of that event. 
This ``loose'' version of STP   captures an intuitive idea of reasoning by cases, which is considered by Savage as a general principle for rational decision making, ``I,'' he continues, ``know of no other extralogical principle governing decisions that finds such ready acceptance.''  In the light of this principle of rationality  we propose  the following generalization of STP:
 
\nocite{Savage72} 
\begin{description}
	\item[Generalized Sure-Thing Principle] {\it If} a decision maker makes the same decision in {all} possible situations based on the information she holds in each situation, {\it then} the decision maker shall decide on {that} decision {without} differentiating the information that leads to the decision.
\end{description} 
Formally, let $D$ be a nonempty set  with unspecified domain, call $f: 2^\Omega\to D$ an \emph{informational decision function for} $i$ if for any $S\subseteq \Omega$ the following condition holds:
\begin{equation}\label{eq:info:min condition 3} \tag{GSTP}
			f\big(\I^i(\nu)\big) =d\;\; \text{for all }\nu\in S \quad \implies \quad  f\Big(\bigcup_{\nu\in S}\I^i(\nu)\Big)=d,
\end{equation}
where $f\big(\I^i(\nu)\big)=d$ is a decision made by player $i$ in state $\nu$ based on her information set $\I^i(\nu)$, and $S$ is a set of possible situations. Then \eqref{eq:info:min condition 3} says that if player $i$ makes the same decision $d$ in all possible situations (states) in $S$ then she should decide on $d$ without differentiating the information generated in $S$ (i.e., $\I^i(\nu)$'s where $\nu\in S$).
We may now prove the following:

\thm 
\label{prop:bs:aumann76}  
	Let $\Omega, N, \I^i$ be defined as above and $\o$ be the actual state of the world.  Suppose that, for any $i,j \in N$, 
	\begin{enumerate}
	    \item $\I^i,\I^j$ are divisible,
	    \item $\B^i(\Omega)=\B^j(\Omega)$;
		\item $f$ is an informational decision function for $i$; and
		\item $i$'s decision $d^i$ is common information shared among members of $N$ at $\o$.
	\end{enumerate}
 Then, $d^i=d^j$ for all $i,j\in N$.
\endthm

If $f$ is intended to be a conditional probability of some  event $A$ (i.e., if $f(\cdot)=\Pr(A\mid \cdot)$) and $\I^i(\Omega)=\Omega$ (and hence $\B^i(\Omega)=\emptyset$ by \propref{lem:bs:Info and blind}), then we have the impossibility result of \cite{Aumann76} as a special case of  \thmref{prop:bs:aumann76}. 

The theorem proved here is more general in that there is no reference to the notion of knowledge or that of beliefs. This has a conceptual advantage: the proof exemplifies the reasoning behind the agreeing-to-disagree type of argument which essentially rests on the structures of information distributions among a group, which can be analyzed from a detached viewpoint without having to ask those puzzling questions like how the common knowledge of prior and posterior probabilities is arrived at.  Moreover, unlike Aumann's knowledge model,   we do not excluded the situations that admit doxastic blindspots, which extends the impossibility theorem to a more general setting. 

However, as seen in the theorem, in order to have the convergence of group opinions it is necessary that all the players in $N$ have the same doxastic blindspots.\footnote{
Aumann's original proof uses a knowledge model  which is provably equivalent to a S5 system in modal logic. As a consequence, it is assumed implicitly that there is no blindspot for any of the players in his model.
} 
Counterexamples can be easily sought if this requirement is dropped. But this is a very strong assumption  on the part of players in a group, where the players may enter the decision situation from different angles and hence have different information.  Therefore, in general it is possible for the players to agree to disagree.\footnote{
For further discussion on this point see \cite{Collins96}.
}

\section{Models of beliefs }\label{sec:BS:models}
\subsection{Belief operator} 
In a standard information model, to say that player $i$  \emph{believes} certain event $E$, in symbols  $B^iE$, is for the player to {have the information} about $E$. Given the characterization of the information structure of player $i$ above, we say that player $i$ \emph{believes} certain event $E$ in state $\o$ if $\I^i(\o)\subseteq E$. Formally, a {\it belief operator} $B$ is a set-valued function on $2^\Omega$ satisfying 
\begin{equation}\label{eq:bs:belief function}
	BE=\{\o\in\Omega\mid \I(\o)\subseteq E\};
\end{equation} 
or, equivalently, by \eqref{eq:bs:info function},
\begin{equation}\label{eq:bs:belief function2}
	B E=\{\o\in\Omega\mid \text{for any $\nu\in \Omega$, $\o\leadsto \nu$ implies $\nu\in E$}\}.
\end{equation} 
The following is a list of properties of a belief operator $B$: for any $E,F\subseteq\Omega$, 
	\begin{description}
		\item[N] $B\Omega=\Omega$
		\item[K] $B(E\cup F)\cap B\neg E\subseteq BF$
		\item[D] $BE\subseteq \neg B\neg E$
		\item[4] $BE\subseteq BBE$
		\item[5] $\neg BE\subseteq B\neg BE$
	\end{description}
where $\neg E$ stand for $\Omega- E$. The labels of the above properties echo the names of the corresponding inference rule of necessitation and the axioms in modal logic. 

\thm\label{lem:bf:B operator}
	Let $B$ be a belief operator defined above, then $B$ satisfies {\bf N} and {\bf K}.
\endthm

\thm\label{lem:bf:B operator2}
	Let $B$ be a belief operator as defined in \eqref{eq:bs:belief function}, then
	\begin{enumerate} 
		\item\label{lem:bf:B operator2: 2}
			$B$ satisfies {\bf D}  if and only if the underlying information structure $\I$ is viable; 
		\item\label{lem:bf:B operator2: 4}
			$B$ satisfies {\bf 4}  if and only if  $\I$ is inclusive;  
		\item\label{lem:bf:B operator2: 5}
			$B$ satisfies {\bf 5}  if and only if  $\I$ is mutual.
	\end{enumerate}
\endthm

In the light of  \propref{thm:ck:property structures} and \propref{lem:bf:B operator2}, let us summarize the relations between the properties of a belief operator $B$ and that of its corresponding defining information structure $\I$ \eqref{eq:bs:belief function} and doxastic accessibility relation $\leadsto$ \eqref{eq:bs:belief function2} in \tableref{tbl:bs:b leaststo I}.
\begin{table}[h] 
\caption{}\label{tbl:bs:b leaststo I}
\centering
\begin{tabular}{ c |  l l }
   $B$  & $\leadsto$ & $\I$    \\
\hline 
  {\bf D} & serial  & viable   \\
  {\bf 4} & transitive & inclusive \\
 {\bf 5} & Euclidean & mutual \\
\end{tabular}
\end{table}
\subsection{Degrees of Belief} In  defining a belief operator \eqref{eq:bs:belief function}, an event $E$ is said to be believed by player $i$ if the information $i$ possesses in state $\o$ is {\it completely} included in  $E$. Now let us consider the case where the inclusion is not complete, that is, the information $i$ has at $\o$ is only partially contained in $E$. In this case, the player is uncertain about the occurrence of $E$. What we need is a more fine-grained measure of beliefs. This is usually fulfilled by using probability measures to represent players's partial beliefs. To be more precise, let $\Omega$ be a state space and $\FF$ be an algebra over $\Omega$. Let player's credal states be represented by a set $\P^i$ of probability functions on $\FF$ (indeterminate probabilities). We refer to $\P^i$ as {\it the global probability assignments of player $i$}. This concept corresponds to players' {\it prior} probability assessments in the game-theoretic literature. We however refrain from using this nomenclature for the reasons to be discussed shortly in Section \ref{sec:remarks:09-03-13}.
 
Presumably, in a given state space $\Omega$, the doxastic accessibility relation $\leadsto^i$ among members of $\Omega$ and the probability functions in $\P^i$ over $\Omega$ are different measures of the same epistemic capability of player $i$, namely $i$'s belief structure. There shall be some compatibility requirements that relate these two different modes of modeling. Recall that $\B^i(\Omega)$ stands for the set of blindspots of   $i$ in $\Omega$.  That is to say, for any $\o\in \B^i(\Omega)$,  the existence of $\o$ in $\Omega$ is postulated by  $i$ as a mere logical possibility to which the player has no epistemic access. It is then conceivable that, in the corresponding probabilistic model of beliefs where epistemic possibilities are measured by some probability functions, a doxastic blindspot be assigned with zero probability. Hence we have the following consistency requirement imposed on $i$'s information structure and priors that,\footnote{
For notational convenience,  write $\rho^i(\o)$ for $\rho^i\big(\{\o\}\big)$.
}
\begin{align}\label{eq:bs:blindspot}\tag{B1}
	  \o\in \blind \;\text{ if and only if }\; \rho^i(\o)=0,\; \text{ for all $\rho^i\in \P^i$}.
\end{align}
In other words, $\o$ is a doxastic blindspot for $i$ if and only if  all the priors of $i$ assign probability  0 to $\o$.\footnote{
To avoid the complications from regularity (cf. \cite{Hajek13} and \cite{Spohn12a}), throughout this paper it is assumed that $|\Omega|<\infty$ and $\FF=2^\Omega$.
}

By \propref{lem:bs:Info and blind}, \eqref{eq:bs:blindspot} implies that, for any $\o\in \I^i(\Omega)$, $\rho^i(\o)=0$ for all $\rho^i\in \P^i$, that is, $\o$ is $\leadsto^i$-accessible from some state if and only if $\rho^i(\o)>0$ for all $\rho^i\in \P^i$. Condition \eqref{eq:bs:blindspot} characterizes a global relationship between $\P^i$ and  $\I^i(\Omega)$ (or $\B^i(\Omega)$); the next question is, at the local level, what is the relationship between $\P^i$ and each $\I^i(\o)$, $\o\in \Omega$. We discuss this in the context of epistemic games. 

\section{Beliefs in epistemic game}\label{sec:BS:emodels}

\subsection{Epistemic extension} 
In this paper, we consider models for games in strategic form augmented by type structure.
Formally, 
let $\Gamma=\brkt{N,\{A^i\},\{u^i\}}$ be a (finite) strategic game, 
where $N$ is a finite group of players and,  for each player $i\in N$,  $A^i$ is a nonempty set of {\it actions} from which player $i$ choose to act.  Let $\A=  \prod_{i \in N} A^i$ be the product space of all action spaces, each member of which forms an {\it action profile}, that is, for any $a\in \A$,  $a=(\xn{a})$ where $a_i\in A^i$.  We refer to a real function $u^i:\A\to \Real$ as the {\it payoff} function of $i$. The interpretation is that if the players in $N$ adopt respectively the actions described in the profile $a\in \A$ then $u^i(a)$ is the {\it payoff} (or the {\it outcome}) for $i$.
Following the  convention in game theory, we often write an action profile $a=(\xn{a})\in \A$ as $(a^i,a^{-i})$ where $a^i$ is the $i$th coordinate of $a$, denoted also by $(a)_i$, i.e., the action adopted by player $i$ in $a$, and $a^{-i}$ is the action profile of $a$ of the players other than $i$. Write $A^{-i}$ for  $\prod_{j\neq i}A^j$, hence, for any $a\in \A$, we have $a^i\in A^i$ and $a^{-i}\in A^{-i}$.  

Further,  let $T^i=\Delta (A^{-i}) $ denote the set of the probability measures defined on $A^{-i}$. We refer to a member $t^i$ of $T^i$ as a {\it type} of player $i$, which characterizes $i$'s belief about the other players' behavior.\footnote{
	The current definition of types is altered from \cite{Aumann95a} which in turn is inspired by \cite{Harsanyi68} types in describing games with incomplete information, see   \cite{Perea13}. The difference is that	we do not include in the present model an explicit representation of a hierarchy of beliefs (cf. \cite{Brandenburger93}); what we call types conrresponds to the concept of {\it conjecture} in \citeauthor{Aumann95a}'s terminology which is considered in the present paper as a constitutive aspect of $i$'s belief. In principle, there is a degree of freedom in generating types. The current definition of types can be easily extended to include other players' types. To be more precise,  a type of player $i$ can be defined as a probability function  on $ A^{-i}\times T^{-i}$. The only restricting requirement that is emphasized in the current paper is not to define type on the players' own actions. And it it is not difficult to see that \thmref{thm:bs:type accessibility 89} and \corref{thm:BS:coherent} below still hold under modified types.
} 
Let $\T=\prod_{i\in N}T^i$ be the type space of $N$, and hence, for any $t=(\xn{t})\in \T$, $t$ provides a {\it type profile} of group $N$. Similarly,  write $T^{-i}$ for  $\prod_{j\neq i}T^j$, then, for any $t=(t^i,t^{-i})\in \T$, $t^i\in T^i$ and $t^{-i}\in T^{-i}$. .

\defn 
	An \emph{epistemic extension} of  strategic game $\Gamma$ is a structure of the form $\Gamma'=\brkt{N,\{A^i\},\{u^i\},\{T^i\}}$. 
\enddefn

\subsection{Belief structures in epistemic games}  Let $\Omega=\A\times \T$, then, for each $\o\in\Omega$, $\o$ is in the form of 
\begin{equation}\label{eq:bs:type state space}
	\o=(a,t)=(a^i,a^{-i};t^i,t^{-i}),
\end{equation}
We refer to each $\o$ as a {\it state-description} of the given game situation which includes a profile of actions and a profile of types of all players. For notational purpose, we use $\bbrkt{a^{-i}}$ to denote the set of all  states in $\Omega$ that contain $a^{-i}$, i.e., 
\begin{align}\label{eq:bs:type and prior:b}
	\bbrkt{a^{-i}}=\Big\{\o\in \Omega \;\Big|\; \o=(b^i,b^{-i};s^i,s^{-i})\text{ and } b^{-i}=a^{-i}\Big\}.
\end{align}
Similarly, $\bbrkt{t^i}$ is the set of states that contain  $t^i$.

Now we return to the discussion of the relationship between different belief structures in $\Omega$.  As before, let $\P^i$ be the set of probability functions over $\Omega$ that represents player $i$'s beliefs. 
We take that player $i$'s types shall not be inconsistent with the probabilistic judgements assessed globally, that is, for any  $t^i\in T^i$ and for any $a^{-i}\in A^{-i}$, there is some $\rho^i\in \P^i$ such that, 
\begin{equation}\label{eq:bs:type and prior}\tag{C1}
	t^i(a^{-i})=\rho^i\big(\bbrkt{a^{-i}}\big).
\end{equation}
Further, to relate to player $i$'s doxastic accessibility relation $\leadsto^i$ in $\Omega$, define that, for any $\o=(a^i,a^{-i};t^i,t^{-i})$ and $\o'=(b^i,b^{-i};s^i,s^{-i})$,
\begin{align}\label{eq:bs:type and accessibility}\tag{C2}
	\o\leadsto^i\o' \quad\text{ if and only if}\quad t^i\big(b^{-i}\big)>0.
\end{align}
That is to say,  state $\o'$ is considered by player $i$ as a doxastic possibility in state $\o$  if $i$ assigns  a non-zero probability to $b^{-i}$ contained in $\o'$. 

\thm\label{thm:bs:type accessibility 89}
	Given \eqref{eq:bs:blindspot} and \eqref{eq:bs:type and prior}, the accessibility relation $\leadsto^i$ ($i\in N$) defined in \eqref{eq:bs:type and accessibility}  is (1) serial, (2) transitive, and (3) Euclidean.
\endthm

The following result  is a direct consequence of \thmref{thm:ck:property structures}, \thmref{lem:bf:B operator2} and \thmref{thm:bs:type accessibility 89} (cf. \tableref{tbl:bs:b leaststo I}).

\cor\label{thm:BS:coherent}
	  For any $i\in N$, let $\I^i$ and $B^i$ be defined in terms of $\leadsto^i$ in \eqref{eq:bs:type and accessibility}, then 
	each $\I^i$ is divisible and $B^i$ satisfies {\bf KD45}.
\endcor
It is shown that given the unifying conditions discussed thus far,  different belief structures can be pieced together in terms of inter-definability in the context of epistemic game theory with rounded properties that are commonly attributed to a full belief structure in  logic.


\section{Discussion}\label{sec:remarks:09-03-13}
 
\subsection{Global and local probabilistic assessments}\label{sec:globle vs local:090313} 
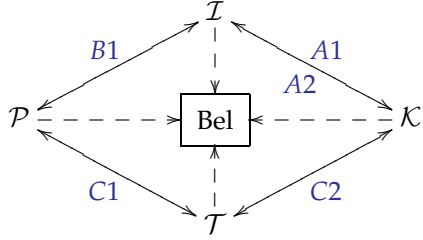
\begin{figure}[h] 
	\begin{displaymath} 
	\xymatrix{
	 &&\I \ar@{<->}[dll]_{\ref{eq:bs:blindspot}}\ar@{-->}[d]  &\\
	\P\ar@{-->}[rr]  &&*++[F-]{\txt{Bel}} &&\K\ar@{-->}[ll] \ar@{<->}[ull]^{\ref{eq:bs:info to leadsto}}_{\ref{eq:bs:info function}}\\
	&& \T\ar@{-->}[u]\ar@{<->}[urr]_{\ref{eq:bs:type and accessibility}} \ar@{<->}[ull]^{\ref{eq:bs:type and prior}}&&
	}
	\end{displaymath} 
	\caption{Unifying belief structures in epistemic games}\label{fg:bs:relations}  
\end{figure}

\figref{fg:bs:relations} is an illustration of the interrelationships between different aspects of the belief structure in epistemic games under the unifying conditions \ref{eq:bs:info function}, \ref{eq:bs:info to leadsto}, \ref{eq:bs:blindspot}, \ref{eq:bs:type and prior}, and \ref{eq:bs:type and accessibility}.  As seen, what links the Kripke structure and the probabilistic structure in the epistemic approach to games is the concept of types \eqref{eq:bs:type and accessibility}. In this paper, the types $\T$ are taken to be decision makers' local probabilistic assessments of other players' behavior, which in turn is considered as a constituent part of the global probabilistic assessments $\P$.  Condition \ref{eq:bs:type and prior} is a consistency constrain that relates these two levels of probabilistic assessments. 
Further, it is worth pointing out that local probabilities provide also a refined measurement of doxastic accessibility. That is, by condition \ref{eq:bs:type and accessibility}, we can now say that $\o'$ is $\leadsto^i$-accessible from $\o$ {\it to degree c} if $t^i(b^{-i})=c$. This accords well with the intuition that some states are ``more'' $\leadsto^i$-accessible than others. 

Let us now return to the discussion of the interpretations of these different levels of probabilistic assignments. The global probability assignments $\P^i$ are often referred to as  {\it prior probabilistic judgments of player $i$} (for models with determinate probability replace $\P^i$ with a single probability function, the conceptual problems raised below apply to both  determinate and  indeterminate priors), which are probability measures defined on the state space $\Omega$. Note that, in the context epistemic game theory, each state-description includes a profile of  actions of all players, see \eqref{eq:bs:type state space}, hence each probability function $\rho\in \P^i$ assigns indirectly a  probability to action of each player including $i$ herself. However, we point out that the assumption of $\P^i$ being the priors of player $i$ is conceptually untenable.\footnote{
The discussion in \ref{sec:pro and act: epi13} is in part modified from Section 3 in \cite{Liu13}.
} 


\subsection{Probability and Act} \label{sec:pro and act: epi13}

The thesis that directly undermines attributing global probability assignments to players as their own priors is this: ``Any adequate quantitative decision model must not explicitly or implicitly contain any subjective probabilities for acts''.\footnote{
	Here we quote the version that is explicitly stated in \cite{Spohn77}. Levi's well-known thesis that ``deliberation crows out prediction" is largely in agreement with this principle, see \cite{Levi96}.
}
In games, the principle says that {\it no subjective probabilities could be assigned by the players to {their own} future actions in a meaningful way}, and hence $\P^i$ cannot be a set of priors of $i$ obtained from the outset.

This doctrine of so-called ``no probability for self-action'' is hinted in Savage's discussion on the ``small world'' semantics where he mentioned in passing that probabilities for acts play no role in individual decision making \citep[pp.82-84]{Savage72}. 
To illustrate, let us use the example Savage provided. Suppose that Jones is torn between either  (i) buying  a sedan, or (ii) buying a convertible, or (iii) canceling altogether the plan of purchase and keeping the money. It is conceivable that, in a simple decision situation, the execution of an action may be solely determined by the relative ranking of the consequences to which the actions lead. That is to say, if Jones prefers the convertible the most, then he shall just proceed to buy a convertible. ``Chance and uncertainty are considered to have nothing to do with the situation.'' 
One might object by suggesting that if Jones likes, say, equally the sedan, he might come to make a decision on his future action of purchasing either a convertible or a sedan by tossing a coin or by utilizing some internal randomizing mechanism (whim, impulse, etc.) to help him  decide, then in this case there seems to be a sense in which one could say that Jones will take an action with such-and-such probability. However the formulation of the objection itself make it plain that the derived probability is essentially about the randomizer in use, which can hardly qualify for the agent's genuine {\it subjective} probability assignment. 

In a more complicated situation where more careful deliberation is required, the agent's actions are further evaluated in terms of their respective consequences under different circumstances, and, according to Savage, only the latter are subject to probabilistic evaluations. Say that Jones has finally reached a decision to buy a convertible because he realized that it's very likely that he will be taking a vacation in Monterey next month, in which case the utility of driving a convertible by the seaside in warm spring breeze will be maximally materialized. Hence, as seen in both  situations, the choices of actions are never deliberated by a decision maker in isolation, they are always placed in a {\it context} in which they are to be evaluated, and the context itself will provide the relevant details (including the circumstances in which an act may take place and the consequences it results in) under which the values of acts are measured and decided. These considerations led Savage to his {\it belief-act-consequence} model, where acts are treated as functions mapping from (act-independent) states to consequences, but {\it not directly as the subjects of uncertainty}.
 
Perhaps, part of the difficulties in incorporating the concept of all-inclusive states in a game-theoretic model is originated from an unquantified reading of Bayesianism. ``According to the Bayesian view, subjective probabilities should be assignable to {\it every prospect}, including that of players choosing certain strategies in certain games'' (\cite{Aumann87} p.2, my emphasis). Here, it is not entirely clear as to which subjective Bayesian theory is under consideration, yet it had been made explicit that each player is assumed to ``conform to the Savage theory.'' However, as we have discussed a few lines above, any assignment of the players' subjective probabilities to their own acts/strategies actually falls outside the scope of the decision model put forward by Savage. 

Yet even if we grant a more general reading on Bayesianism in the respect that subjective probabilities are manifested in their betting interpretation it will soon be clear that no non-trivial probabilities can be assigned to one's own future acts. To see this, suppose that Jones is faced with choices of either going to an Italian restaurant or to a French restaurant for dinner. In an attempt to elicit his subjective probabilities assigned on the two possible future actions, Jones is offered a bet with payoffs as follows
\begin{description}
	\item[i] win $X$ if Jones goes to an Italian restaurant,
	\item[ii] nothing if Jones goes to a French restaurant.
\end{description}
(here it is assumed that the monetary reward of $X$ is not too significant to take effect on Jones' choice of restaurant yet it is not too small to be easily ignored by Jones either). Now, suppose that Jones' subjective probability for his going to an Italian restaurant is $p$, then he shall be willing to pay a fee of $pX$ to accept the bet in exchange of a reward of $X$ on the event that he indeed is having Italian food for dinner. So far the example accords well with the standard betting interpretation of subjective probabilities, but story will soon turn once we notice that Jones' willingness to accept the bet of his going to an Italian restaurant at a cost of $pX>0$ implies that he will be going to an Italian restaurant {\it for sure}! For, otherwise, it would be extremely unwise for Jones to {\it knowingly} pay a fee of $pX$ but actually go to a French restaurant while gaining nothing from the bet he paid for, which can easily be avoided by simply rejecting the bet. And this is true for any $0<p\le 1$. Further, if $p=0$ then Jones will be going to a French restaurant  {\it for sure}. It follows that the betting rate upon which Jones is willing to pay a fair price for his acts collapses into 1 or 0.\footnote{
See \cite{Spohn77} p.115 and  \cite{Levi89} p.32 for relevant discussions on this point.
}  
In another word, personal probabilities tend to be ``gappy'' when it comes to the agents' own actions. 

On a further note, it would  not  be very constructive for us to just bite the bullet and admit that probabilities can only be employed in a ``take it or leave it'' fashion when it comes to self-actions, while continuing to treat $\P^i$ as a set of priors of player $i$, especially when there is an inserted assumption of $\P^i$ being commonly shared among all players (common priors). For it is plain that the latter renders that there is no prior that is not trivial.    

\subsection{A dilemma} The distinction between global and local probability assessments can also be characterized in terms of the asymmetry of different viewpoints, a recurring issue in the foundations of game theory. That is, in laying out different parameters of a game-theoretic model, some aspects of the model are prescribed from the stance of an external observer (theorist's third-person perspective) and some are characteristic of the players being modeled (player's first-person viewpoint). The mismatch between these different levels of viewpoint is the source of many mind-boggling puzzles in game theory.
The often asked question is why a first-person viewpoint cannot be leveled with that of a third-person? Or, in another word, how is it that the players themselves cannot be the theorist? As a matter of fact, many important results presuppose a convergence of the two epistemic perspectives. However, the discussion we made above gives a not-all-favorable answer to these questions, it points out that within the Bayesian framework how the gap created by the asymmetry of different viewpoints {\it cannot} be filled.
 

\section*{Appendix}
 
\proof[Proof of Proposition \ref{lem:bf:divisible property}]
	Let $\o,\o'\in \Omega$, then, by viability, $\I(\o)$ and $\I(\o')$ are non-empty. If $\I(\o)\cap\I(\o')=\emptyset$ then we are done. Otherwise, let $\nu'\in\I(\o)\cap\I(\o')$ and $\nu\in\I(\o)$, from the former we get  $\nu'\in \I(\o)$ and hence $\I(\nu')\subseteq \I(\o')$ by inclusion. Then $\nu\in\I(\o)$ and  $\nu'\in \I(\o)$  implies $\nu\in\I(\nu')$  via mutuality,  hence $\nu\in\I(\o')$. This is shows that $\I(\o)\subseteq   \I(\o')$. Similarly, $\I(\o')\subseteq \I(\o)$. Together, we have $\I(\o)=\I(\o')$. 
\qed 

\proof[Proof of Proposition \ref{lem:bs:Info and i j}] 
	By \defnref{defn:bs:divisible}, $\I^i$ and $\I^j$ are divisible if and only if they are viable, inclusive, and mutual.
	\begin{enumerate}
		\item  If $\o$ is a blindspot it is trivially true that $\o\notin\I^i(\o)$ by \defnref{defn:bs:blindspots} . It remains to  show the ``only if'' direction. Suppose, for contradiction, that $\o\notin\B^i(\Omega)$, then by definition this implies, there is some $\nu\in \Omega$ for which $\o\in \I^i(\nu)$, from which we get $\I^i(\o)\subseteq \I^i(\nu)$ via inclusion. On the other other hand, by viability, $\I^i(\o)$ is non-empty, then let $\o'\in \I^i(\o)$, hence  $\o'\in \I^i(\nu)$. From $\o\in \I^i(\nu)$ and $\o'\in \I^i(\nu)$ we conclude that  $\o\in\I^i(\o')$ and $\o'\in\I^i(\o)$ via mutuality. The latter also implies that  $\I^i(\o')\subseteq\I^i(\o)$, and hence $\o\in \I^i(\o)$, a contradiction. Thus, if $\I^i$ is divisible, then $\o\notin \I^i(\o)$ implies that $\o\in \B^i(\Omega)$.
		\item 
			Suppose, to the contrary, that $\nu\not\in \I^i(\nu)$, then, by \eqref{lem:bs:Info and i j:1}, $\nu\in \B^i(\Omega)$, but this contradicts the assumption that $\nu\in \I^i(\o)$, and hence $\nu\in \I^i(\nu)$. This yields $\nu\in \I^i(\nu)\cap\I^i(\o)$, then, by \propref{lem:bf:divisible property}, $\I^i(\nu)=\I^i(\o)$.
	
		\item For any $\nu\in \I^i(\Omega)$, we have $\nu\in \I^j(\Omega)$. The latter implies that there is some $\o'\in \Omega$ for which   $\nu\in \I^j(\o')$. Then, by \eqref{lem:bs:Info and i j:2}, $\nu\in \I^j(\nu)$.		\qed
	\end{enumerate}
	
\proof[Proof of Proposition \ref{lem:info:basics PN pr}] 
	(1) and (2) are immediate from the definitions above. To see (3), note that  
		\begin{equation}\label{eq:bs:union property}
			\I^i\big(\I^N(\o)\big)=\bigcup_{\nu\in\I^N(\o)} \I^i(\nu).
		\end{equation}
	Now suppose, to the contrary, that there is some $\delta$ such that $\delta\in\bigcup_{\nu\in\I^N(\o)} \I^i(\nu)$ but $\delta\notin \I^N(\o)$. From the latter we conclude that $\delta$ is not $\leadsto^N$-accessible from $\o$;  the former, on the other hand, implies that there is some $\nu\in \I^N(\o)$ such that $\delta\in\I^i(\nu) $, from which we get that $\delta$ can be reached from $\o$ (first to $\nu$ via $\leadsto^N$ and then to $\delta$ through $\leadsto^i$), and hence is $\leadsto^N$-accessible from $\o$, a contradiction.  
	
	For \eqref{lem:info:basics PN pr:4}, it is  sufficient to show that 
	$\I^N(\o)\subseteq   \I^i(\I^N(\o))$ under the assumption that $\I^i(\Omega)= \I^j(\Omega)$ for all $i,j\in N$. Let $\nu\in \I^N(\o)$, then, by \eqref{lem:info:basics PN pr:1}, there shall be a sequence $i_1,i_2,\ldots,i_k\in N$ and a sequence  $\o_0,\ldots,\o_{k-1},\o_k\in \I^N(\o)$ with $\o_0=\o$ and $\o_k=\nu$ such that  $\nu\in \I^{i_k}(\o_{k-1})$. The latter implies that $\nu\in \I^{i_k}(\Omega)$, then, by \propref{lem:bs:Info and i j}\eqref{lem:bs:Info and i j:3}, $\nu\in \I^i(\nu)$, and hence $\nu\in \bigcup_{\nu\in\I^N(\o)} \I^i(\nu)$ in \eqref{eq:bs:union property}. This completes the proof of the lemma.
\qed

\proof[Proof of \thmref{prop:bs:aumann76}] 
	For any $i\in N$, consider the event
		\begin{align}\label{eq:ck:gen11}
			E^i&=\Big\{\nu\in\Omega\mid f\big(\I^i(\nu)\big) =d^i\Big\}, \quad d^i\in D
		\end{align}
	where $E^i$ is the set of possible states in which $f$ yield $d^i$ given player $i$'s information  $ \I^i(\nu)$ at $\nu$. It is plain that $\o\in E^i$ for all $i\in N$. By \defnref{defn:infomodel:cevent}, the assumption that  $d^i$'s are common information at $\o$ amounts to
	  	\begin{align}\label{eq:ck:gen21}
		\I^N(\o)\subseteq\bigcap_{i\in N}E^i.
		\end{align}	
	Then,  $\nu\in \I^N(\o)$ implies $\nu\in E^i$ via \eqref{eq:ck:gen21}, and hence, by \eqref{eq:ck:gen11},   
	\begin{align}\label{eq:ck:gen493}
		 f\big(\I^i(\nu)\big)  = d^i.
	\end{align}
	Note that each $\I^i$ is assumed to be divisible, this implies, by \lemref{lem:info:basics PN pr}\eqref{lem:info:basics PN pr:4}, that, for each $i\in N$,
	\begin{align}\label{eq:ck:gen4931}
		\I^N(\o)=\bigcup_{\nu\in \I^N(\o)} \I^i(\nu).
	\end{align}
	 Finally, since $f$ is an informational decision function for each $i$,  apply \eqref{eq:info:min condition 3} to \eqref{eq:ck:gen493} and \eqref{eq:ck:gen4931}, we get $f\big(\I^N(\o)\big) = d^i$.
	Therefore $d^i=d^j$ for all $i,j\in N$.
\qed

\proof[Proof of \thmref{lem:bf:B operator}] 
	By \eqref{eq:bs:belief function}, for any $\o \in\Omega$, $\I(\o)\subseteq\Omega$, hence $B \Omega =\{\o \in\Omega\mid \I(\o) \subseteq \Omega\}=\Omega$. And, for any $\o \in\Omega$, $\I(\o)\subseteq E\cup F$ and $\I(\o)\subseteq \neg E$ imply that $\I(\o)\subseteq F$. Thus, {\bf N} and {\bf K} hold for $B$.
\qed
\proof[Proof of \thmref{lem:bf:B operator2}] 
	We prove only the first claim, \eqref{lem:bf:B operator2: 4} and \eqref{lem:bf:B operator2: 5} can be shown similarly. Assume that $B$ satisfies {\bf D} we show that, for any $\o\in \Omega$, there is some $\o'$ such that $\o\leadsto\o'$, or equivalently, $\I(\o)\neq\emptyset$. Suppose, to the contrary, that $\I(\o)=\emptyset$ for some $\o\in \Omega$, then we have $\o\in B\emptyset\neq\emptyset$. But, in {\bf D}, let $E=\emptyset$, we have $B\emptyset\subseteq\neg B\neg\emptyset=\neg B\Omega=\neg\Omega=\emptyset$ via \lemref{lem:bf:B operator}, a contradiction.
	
	Conversely, suppose that $\I(\o)\neq\emptyset$ for all $\o\in \Omega$, then, for any $E$,  $\o\in BE$ implies $\emptyset\neq\I(\o)\subseteq  E$, and hence $\I(\o)\nsubseteq \neg E$. Then, by definition, $\o\notin B\neg E$, thus $\o\in \neg B\neg E$. This shows, $BE\subseteq\neg B\neg E$. 
\qed


\proof[Proof of \thmref{thm:bs:type accessibility 89}]
	Let
		\begin{align*}
		\o&=(a^i,a^{-i};t^i,t^{-i}),\\
		\o'&=(b^i,b^{-i};s^i,s^{-i}),\\
		\o''&=(c^i,c^{-i};u^i,u^{-i}).
         \end{align*} 
	\begin{enumerate}[(1)]
	\item Suppose that, for $\o$, there is no $\o'$ such that $\o\leadsto^i\o'$, this implies, by \eqref{eq:bs:type and accessibility}, that, for any $b^{-i}\in A^{-i}$, $t^i(b^{-i})=0$, and hence $\sum_{b^{-i}\in A^{-i}} t^i(b^{-i})=0$. But this contradicts the assumption that $t^i$ is probability function on $A^{-i}$. Thus, for any $\o\in \Omega$, there exists some $\o'$ such that $\o\leadsto^i\o'$.
	\item Assume that $\o\leadsto^i\o'$ and $\o'\leadsto^i\o''$, then,  by \eqref{eq:bs:type and accessibility}, 
			\begin{align}\label{eq:bs:lem type lead prop}
				 t^i(b^{-i})>0; \quad \text{ and }\quad  s^i(c^{-i})>0.
			\end{align}
		We show $t^i(c^{-i})>0$, i.e., $\o\leadsto^i\o''$.  Suppose that this is not the case, then $t^i(c^{-i})=0$, hence, by \eqref{eq:bs:type and prior}, there is some $\rho^i\in \P^i$ for which $t^i(c^{-i})=\rho^i(\llbracket c^{-i}\rrbracket)=0$. This implies that $\bbrkt{c^{-i}}$ is a set of blindspots via \eqref{eq:bs:blindspot}. On the other hand, from $s^i(c^{-i})>0$ we conclude that there is some $\sigma^i\in \P^i$  such that $s^i(c^{-i})=\sigma^i\big(\bbrkt{c^{-i}}\big)>0$. But if $\bbrkt{c^{-i}}$ is a set of blindspots then, by \eqref{eq:bs:blindspot}, $\sigma^i\big(\bbrkt{c^{-i}}\big)=0$, a contradiction. Hence, $\leadsto^i$ is transitive.
		
	\item Finally, assume that $\o\leadsto^i \o'$ and $\o\leadsto^i\o''$, we show that $\o'\leadsto^i\o''$. The assumption implies that $t^i(b^{-i})>0$ and $t^i(c^{-i})>0$, and hence, for some $\rho^i\in \P^i$, $t^i(c^{-i})=\rho^i(\llbracket c^{-i}\rrbracket)>0$.  Now suppose, to the contrary, that $\o'\not\leadsto^i\o''$, that is,  $s^i(c^{-i})=0$. The latter implies, via \eqref{eq:bs:type and prior} and \eqref{eq:bs:blindspot}, that $ \bbrkt{c^{-i}}$ is a set of blindspots, and hence $\rho^i(\llbracket c^{-i}\rrbracket)=0$, a contradiction. Therefore, $\leadsto^i$ is Euclidean.\qed
	\end{enumerate}

\singlespace
\bibliographystyle{chicago}

\end{document}